\newcommand{\be}{\begin{equation}}\newcommand{\ee}{\end{equation}}
\newcommand{\bea}{\begin{eqnarray}}\newcommand{\eea}{\end{eqnarray}}
\newcommand{\p}[1]{(\ref{#1})}
\newcommand\T{\theta_{12}}
\newcommand\Tb{{\bar\theta}_{12}}
\newcommand\Z{Z_{12}}
\newcommand\D{{\cal D}}
\newcommand\Db{\overline{\cal D}}
\documentstyle[12pt]{article}
\begin{document}
\renewcommand{\thefootnote}{\fnsymbol{footnote}}
\thispagestyle{empty}
\hfill hep-th/9504084 \\

\hfill JINR E2-95-172 \vspace{2cm}\\
\begin{center}
{\large\bf THE MINIMAL N=2 SUPEREXTENSION OF THE NLS EQUATION}
\vspace{1.5cm} \\
S. Krivonos\footnote{E-mail: krivonos@cv.jinr.dubna.su}
and A. Sorin \footnote{E-mail: sorin@cv.jinr.dubna.su}
\vspace{1.0cm} \\
{\it JINR--Bogoliubov Laboratory of Theoretical Physics,
141980 Dubna, Moscow Region, Russia} \vspace{1.5cm}
\end{center}
\noindent{\bf Abstract.}
We show that the well known $N=1$ NLS equation possesses $N=2$
supersymmetry and thus it is actually the $N=2$ NLS equation.
This supersymmetry is hidden in terms of the commonly used $N=1$ superfields
but it becomes manifest after passing to the $N=2$ ones.
In terms of the new defined variables the second Hamiltonian structure of the
supersymmetric NLS equation coincides with the $N=2$ superconformal algebra
and the $N=2$ NLS equation belongs to the $N=2$ $a=4$ KdV hierarchy.
We propose the KP-like Lax operator in terms of the $N=2$ superfields which
reproduces all the conserved currents for the corresponding hierarchy.
\vspace{1.5cm} \\
\begin{center}
{\it Submitted to Physics Letters B}
\end{center}
\vfill
\begin{center}
Dubna 1995
\end{center}
\newpage
\renewcommand{\thefootnote}{\arabic{footnote}}
\setcounter{footnote}0
\setcounter{equation}0
\noindent{\bf Introduction.}
Recently, it has been realized [1-4] that  many integrable two dimensional
equations can be supersymmetrized by considering the supersymmetric
extensions of their second Hamiltonian structures.
In particular,
the $N=2$ supersymmetric Boussinesq equation [2] and $N=3,4$ KdV equations
[3,4] have been constructed along this line  starting from the classical
$N=2$ $W_3$ algebra and $N=3,4$ superconformal algebras, respectively.

In contrast, the known $N=2$ supersymmetric extensions of the nonlinear
Schr\"{o}dinger equation (NLS) [5,6] have been constructed in a different way
starting from some {\it ad hoc} assumptions about superfield content of
the theory.
The key problem in the
$N=2$ supersymmetrization of the NLS equation [5,6] is the lack of the
$N=2$ superextension of its second Hamiltonian structure
which is connected in the bosonic case with
the reduced $sl(2)$ Kac-Moody algebra [5].

The idea of our construction of the $N=2$ supersymmetric extension of the
NLS equation comes from a disguised form of the bosonic NLS equation [7,8]
whose second Hamiltonian structure is the $U(1)$ Kac-Moody extension of
the Virasoro algebra. It is more or less evident that the simplest
$N=2$ supersymmetric extension of this second Hamiltonian structure
is the $N=2$ superconformal algebra.

In this letter,  we will explore this idea by explicit constructing of the
minimal $N=2$ supersymmetric NLS equation containing two bosonic and
two fermionic fields (i.e. twice as small as in [5,6]).
Moreover, as we will show
later, this equation can be connected with the $N=1$ NLS equation [9,10]
through the Miura-like transformation.
We propose the Lax operator in terms of the $N=2$ superfield
and discuss some possible generalizations to higher supersymmetry.
We also demonstrate that the $N=1$ NLS equation possesses the hidden
global $N=2$ supersymmetry and it can be rewritten in terms
of the $N=2$ chiral--anti-chiral superfields. \vspace{1cm}

\noindent{\bf Minimal N=2 super NLS equation.}
Let us start with some brief recalling of the salient features of the
description of the
NLS equation within the multi-field representation of the KP hierarchy [7,8].

The NLS hierarchy can be defined via the  Lax operator [7,8]
\be
L= \partial +R\frac{1}{\partial - S}  \label{lax1}
\ee
and the flows
\be
\frac{\partial L}{\partial t} = \left[ L^n_{+},L \right] \;,
\ee
where $\left\{ + \right\}$ denotes the purely differential part of the
{\it n}th power of the Lax operator.
Starting from this Lax operator we can define the first Hamiltonian
structure\footnote{We use the OPE technique instead of the Poisson
brackets.}
\be
R(z_1)S(z_2) = \frac{1}{z_{12}^2},
\ee
with the Hamiltonian $H_3$
\be
H_3=\int \! dx (R^2+RS^2+R'S),
\ee
and the second ones
\bea
R(z_1)R(z_2) & = & \frac{2R(z_2)}{z_{12}} +\frac{R'(z_2)}{z_{12}^2} \; ,
                          \nonumber \\
S(z_1)S(z_2) & = & -\frac{2}{z_{12}^2} \; , \nonumber \\
R(z_1)S(z_2) & = & \frac{2}{z_{12}^3} + \frac{S(z_2)}{z_{12}} +
          \frac{S'(z_2)}{z_{12}^2}  \label{SHS1}
\eea
with Hamiltonian $H_2$
\be
H_2=\int \! dx RS .   \label{H2}
\ee
for the first nontrivial flow equations [7,8]
\be
\frac{\partial S}{\partial t} = -S''+2S'S+2R' \quad , \quad
\frac{\partial R}{\partial t} = R''+2(RS)' \quad , \label{nls1}
\ee
which are a disguised form of the NLS equation. Namely, the standard
form of the NLS equation
\be
\frac{\partial u}{\partial t} = u''-u^2v \quad , \quad
\frac{\partial v}{\partial t} = -v''+v^2u \quad , \label{nls0}
\ee
coincides with \p{nls1} after passing to the new fields $S$ and $R$
defined as
\be
R=-\frac{1}{2}uv \quad , \quad S=-\frac{v'}{v} \; . \label{transf1}
\ee
Let us note that the standard complex conjugated
rules for $u,v$ and $t$
\be
u^* = v \; , \; v^* =u \;, \; t^* = -t
\ee
have the following representation in terms of $R$ and $S$:
\be
R^* = R \; , \; S^* = -S - \frac{R'}{R} \; .
\ee
Quite unexpectedly,  the standard conjugated rules for $R$ and $S$
\be
R^* = R - S' \; , \; S^* = -S  \; , \; t^* = -t
\ee
induce the B\"{a}cklund-Schlesinger transformations for the $u,v$ system
\cite{JM}
\be
u^* = v\left( u v -2(\log v)''\right) \; , \; v^* =\frac{1}{v} \;,
\ee
which are useful for solving the NLS equation.

It is easy to recognize that the second Hamiltonian structure \p{SHS1}
for the NLS equation in the form \p{nls1} is an extension of the Virasoro
algebra with the stress-tensor $R$ by the
$U(1)$ current $S$ in the non-primary basis.

This Hamiltonian structure can be immediately extended to the $N=2$
supersymmetric case. Indeed, the $N=2$ bosonic supercurrent $J$
generating
the $N=2$ superconformal algebra (SCA) through the following SOPE's
\be
J(Z_1)J(Z_2) = \frac{c/4}{\Z^2} + \frac{\Tb\Db J(Z_2)}{\Z}-
      \frac{\T\D J(Z_2)}{\Z}+
        \frac{\T\Tb J(Z_2)}{\Z^2}+\frac{\T\Tb J(Z_2)'}{\Z} \; , \label{N2SCA}
\ee
where
\begin{equation}
Z=(z,\theta , \bar\theta ), \;
\T=\theta_1-\theta_2 , \; \Tb=\bar\theta_1-\bar\theta_2, \;
  \Z=z_1-z_2+\frac{1}{2}\left( \theta_1\bar\theta_2
       -\theta_2\bar\theta_1 \right)
\end{equation}
and $\D,\Db$ are the spinor covariant derivatives
\begin{equation}\label{3}
\D=\frac{\partial}{\partial\theta}
 -\frac{1}{2}\bar\theta\frac{\partial}{\partial z} \quad , \quad
\Db=\frac{\partial}{\partial\bar\theta}
 -\frac{1}{2}\theta\frac{\partial}{\partial z}
\end{equation}
$$
\left\{\D,\Db \right\}= -\frac{\partial}{\partial z} \quad , \quad
\left\{\D,\D \right\} = \left\{\Db,\Db \right\}= 0,
$$
contains in the bosonic sector (after putting all fermions equal to zero)
just the $U(1)$ extension of the Virasoro algebra, i.e. the second
Hamiltonian structure of the $N=0$ NLS equation \p{SHS1} (in the primary
basis).
Therefore, it is natural to suppose that the second Hamiltonian
structure of the $N=2$ superextension of the NLS equation we are looking for
coincides with $N=2$ SCA \p{N2SCA}. Due to the dimensionality of the
Hamiltonian $H_2$ \p{H2} in the bosonic case $(cm^{-2})$, there is a
unique candidate to be its $N=2$ superextension:
\be
sH_{2}= \int \! dz d\theta d\bar{\theta} J(Z)J(Z) \quad , \label{sH2}
\ee
and the equation of motion is easy to compute
\be
\frac{\partial J}{\partial t} = -\frac{c}{4}\left[ \D , \Db \right] J'+
           4 J'J . \label{N2NLS}
\ee
It will be of importance that the central charge $c$ is nonzero in
\p{N2NLS}, but its concrete value is nonessential and can be changed by
rescaling of $J$ and $t$.
Henceforth, we fix the central charge  equal to $c=-4$ and
call  equation \p{N2NLS} the minimal $N=2$ supersymmetric NLS
equation.

Before going further and to avoid any failure to understand, let us clarify
some points.

First of all  eq. \p{N2NLS} is not a new one because it belongs
to the integrable $N=2$ KdV hierarchy with the parameter
$a=4$ [12]. So, the
main question we have to answer immediately is why we call this equation
the $N=2$ supersymmetric extension of the NLS one? To answer, let us
pass from the superfield equation \p{N2NLS} to the component ones
\bea
\frac{\partial S}{\partial t} & = & -S''+2S'S+2R' \quad , \nonumber \\
\frac{\partial R}{\partial t} & = & R''+2(RS)' + 8 (\xi\bar{\xi})' \quad ,
                           \nonumber \\
\frac{\partial \xi}{\partial t} & = & \xi''+ 2(\xi S)' \quad , \nonumber \\
\frac{\partial \bar{\xi}}{\partial t} & = & -\bar{\xi}''+2 (\bar{\xi} S)',
\label{nls2c}
\eea
where the component currents are defined as follows:
\be
S=2J| \; , \; R=\left( \left[ \D,\Db \right] J  + J'\right) | \; , \;
\xi=\D J| \; , \; \bar{\xi} = \Db J |
\ee
and the sign $|$ means putting $\theta , \bar{\theta}$ equal to zero.

Now it is the matter of calculations to see that  equations \p{nls2c}
are of the disguised form of the $N=1$ NLS equation [9,10].
Indeed, the $N=1$ NLS equation has the following form:
\bea
\frac{\partial u}{\partial t} & = & u''-u^2v - u(\psi\bar{\psi}'-
                   \psi'\bar{\psi})- u'\psi\bar{\psi} \quad , \nonumber \\
\frac{\partial v}{\partial t} & = & -v''+uv^2 + v(\psi\bar{\psi}'-
                   \psi'\bar{\psi})- v'\psi\bar{\psi} \quad , \nonumber \\
\frac{\partial \psi}{\partial t} & = & \psi''- \psi u v +
         \psi\psi'\bar{\psi}  \quad , \nonumber \\
\frac{\partial \bar{\psi}}{\partial t} & = & -\bar{\psi}''+ \bar{\psi}uv-
         \psi\bar{\psi}\bar{\psi}' .       \label{nls1c}
\eea
After passing to the new fields defined as
\bea
S & = & -\frac{1}{2}\psi\bar{\psi}-\frac{v'}{v} ,\nonumber \\
R & = & -\frac{1}{2}\psi\bar{\psi}'-\frac{1}{2}u v , \nonumber \\
\xi & = & \frac{1}{4}\psi v , \nonumber \\
\bar{\xi} & = & -\frac{1}{4}\bar{\psi} u +\frac{\bar{\psi}''}{2v}-
     \frac{\bar{\psi}'v'}{2v^2} \label{miura}
\eea
we recover  equations \p{nls2c}.

Thus, we proved that our $N=2$ NLS equation is nothing else but the
disguised
form of the $N=1$ NLS one and possesses the manifest $N=2$ superconformal
symmetry. Let us stress that the transformations
\p{miura} are  Miura-like ones. So,  equations \p{nls2c} are equivalent
to the $N=1$ NLS equations \p{nls1c} in the same sense as the KdV equation
is equivalent to the mKdV one.

Now, recognizing the second Hamiltonian structure of the $N=2$ NLS equation
as $N=2$ SCA it is natural to generalize it for  higher supersymmetries,
considering the $N=3,4$ SCA's. The corresponding $N=3,4$ superextensions
of the NLS equation will be the first non trivial equations in the $N=3,4$
super KdV hierarchies [3,4]. The detailed calculations for these cases
would be presented elsewhere.

Let us finish this letter with two comments.

First, it can be verified that the Lax operator for our $N=2$ NLS equation
has the following interesting form:
\be
L= \hat{L}+L^{\frac{1}{2}} \quad , \label{lax}
\ee
where $\hat{L}$ and $L^{\frac{1}{2}}$ are two different square roots from
the Lax operator $L_{\mbox{KdV}}$ for the $a=4$ $N=2$ super KdV equation [12]
\bea
L_{\mbox{KdV}} & = & \partial^2 + 2J\left[ \D , \Db \right]+
       2(\D J)\Db-2(\Db J)\D + \left( \left[ \D , \Db \right]J\right) +J^2 ,
         \nonumber \\
\hat{L} & = & \left[ \D , \Db \right] + J  \label{lkdv}
\eea
and $L^{\frac{1}{2}}$ is a pseudodifferential operator starting with
$\partial $:
\be
L^{\frac{1}{2}}=\partial +
 \left( \frac{1}{2} \left( \left[ \D,\Db \right]J \right)+
 J \left[ \D,\Db \right]+ (\D J) \Db - (\Db J) \D \right) \partial^{-1}+
  \ldots  \quad .
\ee
The corresponding Lax equation has the standard form:
\be
\frac{\partial L}{\partial t} = \left[ L^2_{+},L \right] .
\ee
It is interesting enough that using our Lax operator \p{lax}
all the $N=2$ super KdV hierarchy with the parameter $a=4$ can be obtained
from the following equations:
\be
\frac{\partial L}{\partial t} = \left[ L^n_{+},L \right] \label{hkdv}
\ee
in contrast with [12] where only the equations with even $n$ are reproduced.
Our Lax operator \p{lax} is a natural generalization of the Lax pairs
representation proposed in [13] for each of the $N=2$ super KdV
hierarchy equations and describes  some reduction of the $N=2$ KP hierarchy.

Finally, we would like to stress that the $N=1$ NLS equation \p{nls1c}
can be  rewritten
in terms of the $N=2$ chiral-anti-chiral fermionic superfields
$F(Z),\bar{F}(Z)$  as
\bea
\frac{\partial F}{\partial t} = F'' - F \D( \bar{F}\Db F) , & & \nonumber \\
\frac{\partial \bar{F}}{\partial t} =-{\bar F}'' +
           \bar{F} \Db( F\D \bar{F}) , & & \label{sfeq}\\
\D F = \Db \bar{F} =0 ,\;  F^* = {\bar F} \; . & & \nonumber
\eea
where the components of superfields $F$ and $\bar{F}$ are defined as follows:
\be
u = \Db F| \; , \; v = \D \bar{F}| \; , \;
\psi  = F| \; , \; \bar{\psi} = \bar{F} | \quad .
\ee
In terms of the superfields $F$ and $\bar F$ the transformations \p{miura}
have the following form:
\be
J= -\frac{1}{4} F {\bar F} - \frac{1}{2}\; \frac{ \left( \D {\bar F}\right)'}
            {\D {\bar F}} \; .
\ee
This demonstrates the hidden $N=2$ supersymmetry of the $N=1$ NLS equation
\p{nls1c},
but in this form it is unclear how to promote this global $N=2$ supersymmetry
to the $N=2$ superconformal one which is manifest in terms of the $N=2$
superfield $J(Z)$ \p{N2NLS}.\vspace{1cm}\\

\noindent{\bf Conclusion.}
In this paper, we have shown that the well known $N=1$ NLS equation
possesses $N=2$
supersymmetry and thus it is actually the $N=2$ NLS equation.
This supersymmetry is hidden in terms of the commonly used $N=1$ superfields
but it becomes manifest after passing to the $N=2$ ones. In terms of the new
defined variables the second Hamiltonian structure of the supersymmetric NLS
equation coincides with the $N=2$ superconformal algebra and the $N=2$
NLS equation belongs to the $N=2$ $a=4$ KdV hierarchy.
We also constructed the KP-like Lax operator in terms of the $N=2$ superfields
which reproduces all the conserved currents for the corresponding hierarchy.
The relation
between the $N=2$ NLS and $N=2$ KdV equations can be easily promoted to
higher supersymmetries. We postpone the discussion of the NLS equations with
higher supersymmetries to future publications.

In the forthcoming paper \cite{KST} we will show that the appearance of
the hidden $N=2$ supersymmetry in the $N=1$ NLS equation has a nice geometric
description in the framework of the coset approach. \vspace{1cm} \\

\noindent{\bf Acknowledgments.} It is a pleasure to thank L. Bonora,
E. Ivanov, A. Pashnev and F. Toppan for their interest in this work and
clarifying discussions.

\end{document}